\documentclass[12pt]{article}
\usepackage{epsfig}
\usepackage{psfig}
\begin{document}
\begin{titlepage}
\begin{center}

\vspace{5cm}

  {\Large \bf  Role of Glueballs \\ in Non-Perturbative Quark-Gluon Plasma }
  \vspace{0.50cm}\\
  Nikolai Kochelev$^{a,b}$\footnote{kochelev@theor.jinr.ru},
  Dong-Pil
Min$^a$ \footnote{dpmin@snu.ac.kr} \vspace{0.50cm}\\
{(a) \it Department of Physics and Astronomy, Center for Theoretical Physics, \\Seoul National
University,
  Seoul 151-747, Korea}\\
\vskip 1ex {(b) \it Bogoliubov Laboratory of Theoretical Physics,
Joint Institute for Nuclear Research, Dubna, Moscow region,
141980, Russia} \vskip 1ex
\end{center}
\vskip 0.5cm \centerline{\bf Abstract}
 Discussed is how non-perturbative properties of quark gluon plasma,
  recently discovered in RHIC experiment, can be related to the change of properties
   of  scalar and pseudoscalar glueballs.  We set up a model with the Cornwall-Soni's
   glueball-gluon interaction, which shows that the pseudoscalar glueball becomes
    massless above the critical temperature of deconfinement phase
    transition. This change
     of properties gives rise to the change of sign of the gluon condensate at $T > T_c$.
    We discuss the other physical consequences resulting
    from the
       drastic change of the pseudoscalar glueball mass above the critical temperature.
\vskip 0.3cm \leftline{Pacs: 24.85.+p, 12.38.-t, 12.38.Mh, 12.39.Mk} \leftline{Keywords: quarks, gluons, glueball, plasma, non-perturbative QCD}
\vspace{1cm}
\end{titlepage}
\setcounter{footnote}{0}
\section{Introduction}
The results obtained recently at RHIC suggest the formation of a new phase of nuclear matter,
 i.e. the strongly interacting quark-gluon plasma, in high energy heavy ion collision  \cite{STAR,PHENIX}.
  The origin of such phase might be related to the survival of
some strong non-perturbative QCD
   effects  above  the deconfinement temperature (see, for example,~\cite{shuryakQGP,rho}).
    This conjecture is supported by the lattice result for the pure $SU(3)_c$ theory
     \cite{lattice}, which shows that the gluon condensate changes its sign and remains to
      be large till very high temperature as the system crosses over the deconfinement temperature
      \footnote{In the full QCD there is a shift of temperature where the sign change of the gluon condensate
      is taking place \cite{lattice}.}.
 This behavior of gluon condensate forms a clear contrast to that of quark which vanishes at the
 deconfinement transition \cite{lattice2}. Therefore, it is evident that
  the gluonic effect plays the leading role in the dynamic of quark-gluon plasma (QGP) above
  the deconfinement temperature and may give the clue to the unexpected properties of matter produced at RHIC.
The change of sign of the gluon condensate at the finite temperature may give the influence on the glueball properties,
 because it is well known at zero temperature that the gluon condensate plays the role of fixing the mass scale of
  glueball and therefore its dynamics \cite{forkel,kochelevmin}. It is our goal of this work to investigate
  the influence at finite temperature.

  We anticipate the change of the structure of QCD vacuum at $T>T_c$,
  which may lead to the strong modification of the properties of scalar and pseudoscalar glueballs above
   the deconfinement temperature. Some possible signatures of the change of
    properties are already discussed for the scalar glueball in QGP in the recent
     papers by Vento
    \cite{vento}, but not for the pseudoscalar  glueball. Indeed, glueball
    properties should be strongly modified along with
the change of gluon condensate. In the normal QCD, however, the
role of two types of glueball may not be important because of
their large masses. In this Letter within effective Lagrangian
approach, based on the specific non-perturbative gluon-glueball
interaction, it is suggested that the pseudoscalar glueball can
be much lighter above $T_c$. Therefore its role in such high
temperature QGP could be substantially enhanced. This light
glueball may lead the equation of state, and play the role of the
mediator of the strong interaction between gluons in this
environment, more than what the screening perturbative gluon does.

\section{Glueball-gluon interaction and glueball mass above~$T_c$ }
In  \cite{soni}  Cornwall and Soni proposed a simple form of the effective scalar and pseudoscalar glueball
 interaction with gluons,
 \begin{eqnarray} {\cal
L}_{Ggg}=\frac{b_0}{16\pi<S>}\alpha_sG^a_{\mu\nu}G^a_{\mu\nu}S+\frac{\xi
b_0}{16\pi<S>} \alpha_sG^a_{\mu\nu}\widetilde{G}^a_{\mu\nu}P,
\label{lag1}
\end{eqnarray}
where $G^a_{\mu\nu}$ is gluon field strength and
$\widetilde{G}^a_{\mu\nu}=\epsilon_{\mu\nu\alpha\beta}{G}^a_{\alpha\beta}/2$.
$S$ and $P$ are scalar and pseudoscalar glueball fields, respectively, and
 $b_0=11N_c/3$ for pure $SU(3)_c$, $\xi(\approx 1)$  is the parameter of
  the violation of $S-P$ symmetry \footnote{It can be shown within instanton model
for QCD vacuum \cite{shuryak} that  the effective scalar and
pseudoscalar couplings in Eq.\ref{lag}  should be identical due
to the self-duality of instanton field.}, and
 $<S>$ is the nonvanishing expectation
value of scalar glueball field with respect to the vacuum.
They incorporate the low-energy QCD theorems by identifying correlators with gluonic operators
\begin{eqnarray}
J_S&=&\alpha_sG^a_{\mu\nu}G^a_{\mu\nu},\nonumber\\
J_P&=&\alpha_sG^a_{\mu\nu}\widetilde{G}^a_{\mu\nu},
\label{current}
\end{eqnarray}
Note that $<S>$ is related to the gluon condensate $<g^2G^2>$ with the
strong coupling constant $g$ as follows
\begin{equation}
<S>^2=\frac{b_0}{32\pi^2}<g^2G^2>. \label{LETF}
\end{equation}
This equation can be rewritten in more convenient way with the mass of scalar glueball $M_S$ at zero temperature by low energy theorem. (See  \cite{LET} and references therein.)
\begin{equation}
f_S^2M_S^2\simeq\frac{8}{b_0}<g^2G^2>, \label{LETF1}
\end{equation}
where $f_{S}$ is the residue
\begin{equation}
f_{S}M^2_{S}=<0|J_{S}|S>. \label{res}
\end{equation}
From the comparison of Eq.\ref{LETF} and Eq.\ref{LETF1}  we get the simple
relation
\begin{equation}
<S>\simeq\frac{b_0}{16\pi}f_S.
 \label{s}
\end{equation}
Therefore, Eq.\ref{lag1} can be rewritten for the zero temperature as
\begin{eqnarray} {\cal L}_{Ggg}=\frac{1}{f_S}(\alpha_sG^a_{\mu\nu}G^a_{\mu\nu}S+
\xi\alpha_sG^a_{\mu\nu}\widetilde{G}^a_{\mu\nu}P). \label{lag}
\end{eqnarray}
\begin{figure}[htb]
\centering \centerline{\psfig{file=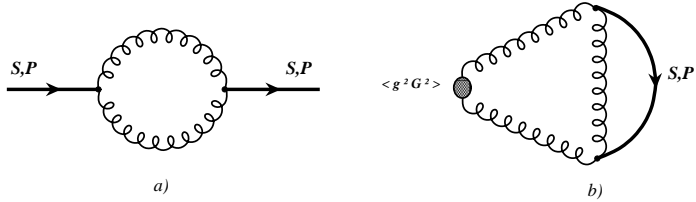,width=10cm,angle=0}} \vspace*{-0.6cm} \caption{ (a) Diagrams for the contribution to the
glueball mass, (b) and for the contribution to gluon condensate. }
\end{figure}
 The effective mass of glueball at $T>T_c$ for pure $SU(3)_c$
can be calculated by considering the contribution of the diagram
presented in Fig.1a to the mass operator $\Pi(p)$ by using
standard methods of finite-temperature field theory
\cite{kapusta}. The result is following
\begin{eqnarray}
\Pi(p)_{S,P}&=&
i\frac{4\alpha^2_s}{\pi^4f^2_{S}}\Bigl[\frac{1}{2}\int_{-i\infty}^{i\infty}dk_0d\vec{k}(D(k_0,\vec{k})
+
D(-k_0,\vec{k}))\nonumber\\&+&\int_{-i\infty+\epsilon}^{i\infty+\epsilon}dk_0d\vec{k}(D(k_0,\vec{k})
+ D(-k_0,\vec{k}))\frac{1}{e^{k_0/T}-1}\Bigr], \label{fin}
\end{eqnarray}
where
\begin{equation}
D(k_0,\vec{k})=\frac{F_{S,P}(k_1,k_2)}{(k^2_1-m^2_g)(k^2_2-m^2_g)}F_{cut}(k^2_1,k^2_2),
\label{mass}
\end{equation}
and a simple form for nonperturbative gluon propagator as a free
propagator  with the effective mass $m_g$ has been used. In
Eq.\ref{mass}, $k_1=p+k$, $k_2=k$, and function $F_{cut}$ provide
the ultraviolet cut-off in the Eucledian space. In
Eq.\ref{mass}, numerator for scalar and pseudoscalar
glueballs can be rewritten as
\begin{eqnarray}
F(k_1,k_2)_S&=&2(k_1.k_2)^2+k_1^2k_2^2\nonumber \\&= &p^2k^2+6k^2(p.k)+2(p.k)^2+3k^4,\nonumber\\
 F(k_1,k_2)_P&=&2\xi^2((k_1.k_2)^2-k_1^2k_2^2)) \nonumber\\
&=&2\xi^2((p.k)^2-p^2k^2). \label{func}
\end{eqnarray}
As usual, we define the effective mass of the glueball as a static
infrared limit of mass operator
\begin{equation}
M_{S,P}^2=\Pi_{S,P}(p_0=0,\vec{p}\rightarrow 0). \label{mass2}
\end{equation}
After Wick rotation to Euclidean space and with assumption about
Gaussian form of cut-off function in this space
\begin{equation}
F^{E}_{cut}(k_1^2,k_2^2)=e^{-\Lambda^2( k_1^2+k_2^2)}, \label{cut}
\end{equation}
the calculation of the integrals in Eq.\ref{mass} leads to
\begin{eqnarray}
M^2_S(T)&\approx&\frac{72\alpha^2_s}{\pi^2\Lambda^4f^2_S}\int_0^\infty
\frac{tdt}{(2+t)^4}e^{-tm^2_g(T)\Lambda^2},\nonumber\\
M^2_P(T)&=&0, \label{mass3}
\end{eqnarray} where we assume that at $T_c<T<2T_c$ the effective mass of gluon
coincides with its  thermal mass \cite{lattice4}
\begin{equation}
m_g=m_g(T)\approx 3T. \label{mg}
\end{equation}
We will neglect the contribution coming from the second term in the
right hand side of  Eq.\ref{fin}, because, in our simple model with heavy
effective gluon thermal mass such contribution should be
suppressed by factor $T^2/m_g^2(T)\approx 1/10 $
 \footnote{To take into account of
this effect is beyond the scope of our accuracy estimations
involved in some values of glueball-gluon couplings, shape of
cut-off function and their possible temperature dependency. Such
contribution, however, could not change our conclusion of
vanishing of pseudoscalar glueball mass above $T_c$.}.

 Most remarkable result of Eq.\ref{mass3} is that the
pseudoscalar glueball mass vanishes at $T>T_c$ due to the
interaction Eq.\ref{lag}. We should point out  that below $T_c$
one can expect that the glueball interaction with gluons has no
such a simple form because confinement forces should also be
included in the consideration. The zero effective mass of the
pseudoscalar glueball in QGP follows from the specific Lorenz
structure of pseudoscalar glueball-gluon interaction and,
therefore, this result does not depend on some particular values
of the our model parameters.  Note that the mass of pseudoscalar
glueball, arising from the interaction Eq.\ref{lag}, is
proportional to so-called topological susceptibility $\chi(T)$
\begin{eqnarray}
M_P^2(T)&=&-\frac{2i}{f^2_S}\int d^4x
<T\alpha_sG^a_{\mu\nu}(x)\widetilde{G}^a_{\mu\nu}(x)\alpha_sG^a_{\mu\nu}(0)\widetilde{G}^a_{\mu\nu}(0)>
\nonumber\\&=&\frac{128\pi^2}{f^2_S}\chi(T), \label{topcharge}
\end{eqnarray}
which vanishes above $T_c$ at the lowest order of our model. Zero
topological susceptibility above deconfinement temperature is in
good  agreement with the recent lattice
 calculation for pure $SU(3)_c$ which shows  a sharp drop of topological susceptibility
   across the deconfinement
   transition \cite{top} and suggests much  simpler topological structure of the strong
    interaction
   above $T_c$ in comparison with the confinement regime.
In contrast to the pseudoscalar glueball, the scalar glueball
remains rather massive even for $T>T_c$. The temperature
dependency of glueball masses is presented in the Fig.2. In the
region $0<T<T_c$ we assume that values of glueball masses are
equal to their zero temperature values, which is consistent with
the observation produced from lattice calculations of very small
change of the gluon condensate in this temperature interval
\cite{lattice}, \cite{lattice2}.
\begin{figure}[htb] \centering
\centerline{\epsfig{file=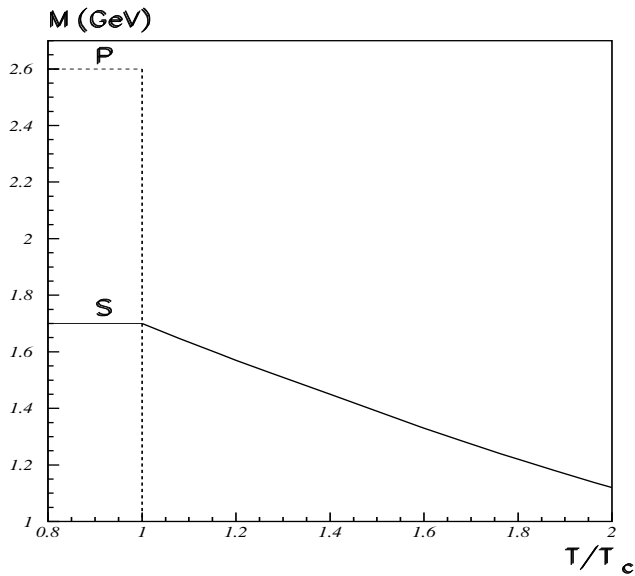,width=10cm,height=8cm,
angle=0}}\ \caption{ The solid (dotted) line is the  temperature
dependence of scalar (pseudoscalar) glueball mass.}
\end{figure}
The value of residue $f_S=0.35$ GeV has been fixed from
Eq.\ref{LETF1} by using $<g^2G^2>\approx 0.5$ GeV$^4$  for the
value of gluon condensate at zero temperature \cite{SVZ} and value
of scalar glueball mass $M_S(0)\approx 1.7$ GeV from quenched
lattice results \cite{lattice3}. We use also the recent lattice
results for thermal mass of gluons above  deconfinement
temperature, $\alpha_s\approx 0.5$ and value of the deconfinement
temperature for pure $SU(3)_c$, $T_c\approx 270$ MeV
\cite{lattice5}. For the cut-off parameter $\Lambda$  in
Euclidean space in  Eq.\ref{cut}, the value $\Lambda\approx
1/M_S(0)$ has been taken \footnote{We assume that $ f_S $ and
$\Lambda$ do not change their values over the deconfinement
transition.}.
 We should
point out that at $T_c<T<2T_c $, the mass of scalar glueball is
large, $M_S(T)>>T$, therefore, one can not expect significant
contribution of this glueball to the bulk properties of QGP in
this range of temperatures. On the other hand, the massless
pseudoscalar glueball should play an important role in the
thermodynamics of QGP at $T>T_c$. Indeed, the simple estimation
shows that the ratio of the scattering amplitudes of gluon-gluon
in the t-channel  exchanged by massless glueball to that by
massive gluon, Eq.\ref{mg}, should be about
$\alpha_sm_g^2(T)/f_s^2>>1$ at $T>T_c$. Therefore, one might
expect the dominance of glueball exchange over screening
perturbative one-gluon exchange at $T>T_c$.

\section{Gluon condensate}

The contribution of glueballs to the gluon condensate  in the
lowest order of the effective coupling constant  is presented in
Fig.1b. The result of calculation is
\begin{equation}
<g^2G^2(T)>_{S,P}=\frac{24\alpha^3_s}{\pi^3f^2_{S}\Lambda^6}\int_0^1dx\int_0^\infty
dt \int_0^\infty dy \Phi_{S,P}(t,x,y), \label{cond1}
\end{equation}
\begin{figure}[htb]
\centering
\centerline{\epsfig{file=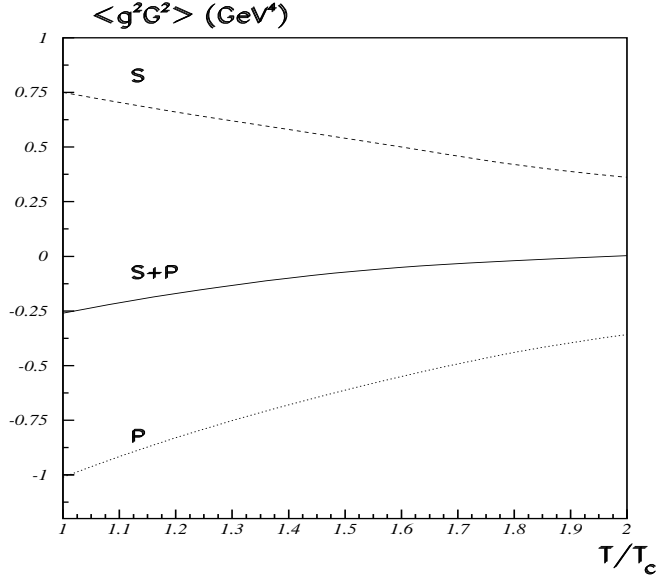,width=10cm,height=8cm,
angle=0}}\ \caption{  The temperature dependency of the gluon
condensate at $T>T_c$. The solid line is total condensate, the
dashed (dotted) line is the scalar (pseudoscalar) glueball
contributions.}
\end{figure}
where
\begin{eqnarray}
\Phi_S(t,x,y)&=&
  \frac{ty^2(1 - x)}{(1+y)^6(t + m^2_g(T)\Lambda^2)}
(3t(1+y)^2 + t^3x^4y^4 +
        5t^2x^2y^2(1+y))\nonumber\\
    &\times & exp[{-\frac{t(1 + y(1 + x) + y^2x(1 - x))}{1+y} - yxM_S(T)^2\Lambda^2 -
        y(1 - x)m_g^2(T)\Lambda^2}]\nonumber\\
\Phi_P(t,x,y)&=&
  -\frac{\xi^2ty^2(1 - x)}{(1+y)^5(t + m^2_g(T)
\Lambda^2)}(3t(1+y) + t^2x^2y^2 )\nonumber\\
    &\times & exp[{-\frac{t(1 + y(1 + x) + y^2x(1 - x))}{1+y} - yxM_P(T)^2\Lambda^2 -
        y(1 - x)m_g^2(T)\Lambda^2]}
\nonumber
\end{eqnarray}
In Fig.3 the dependency of the gluon condensate on temperature for
$T_c<T<2T_c$ is shown. We should mention that the contribution of
the pseudoscalar glueball   to gluon condensate is proportional to
$\xi^2$. Therefore it is quite sensitive to the degree of
violation of the S-P symmetry shown in Eq.\ref{lag}.
 In our numerical estimation we assume that
parameter $\xi\approx 1$  in the light of the Peccei-Quinn (PQ)
mechanism \cite{peccei} (see discussion in \cite{soni}). The
decrease of the $\xi$ value  will lead to the decrease of the
pseudoscalar glueball contribution to the condensate \footnote{
In our model the change of the sign of the total glueball
contribution to the condensate at $T=T_c$ turns out to be
possible only for the large value of that parameter, $\xi>0.86$.
Such large value of $\xi$ for the full QCD could result from the
restoration of $U(1)_A$ symmetry at $T>T_c$ and PQ mechanism
which relates this symmetry to the S-P symmetry.}.
 At zero temperature by using
Eq.\ref{cond1} we get the following values of condensates
\begin{eqnarray}
<g^2G^2(0)>_{total}&=&0.47 GeV^4, \ \ <g^2G^2(0)>_S=1.31 GeV^4,
\nonumber \\
 <g^2 G^2(0)>_P&=&-0.84 GeV^4,
 \label{zeroT}
\end{eqnarray}
where we use the quenching result $M_P=2.6$ GeV for the mass of
pseudoscalar glueball at zero temperature \cite{lattice3}.
 One can see that scalar glueball contribution to gluon
condensate is positive and pseudoscalar glueball contribution is
negative. Total gluon condensate strongly depends on the masses of
glueballs. Note that its  value at zero temperature is taken to be in
agreement with QCD sum rule result \cite{SVZ}. It is evident from
Fig.3 that just above $T_c$ the negative contribution of
massless pseudoscalar glueball to gluon condensate is greater
than the positive contribution coming from massive scalar
glueball and this is a reason of change of the sign of gluon
condensate above the deconfinement temperature in our model
 based on the effective glueball-gluon interaction Eq.\ref{lag}.
Unfortunately the direct comparison of our result with lattice
data \cite{lattice} is not possible due to the necessity of the
 subtraction of the perturbative contribution to
$<g^2G^2(T)>$ from the data \footnote{ We are grateful to S.-H.
Lee for the discussion of this problem.}. We would like to
emphasize that in our  calculation the effective dimensionless
coupling for interaction Eq.\ref{lag}    is  very small,
$g_{eff}=\alpha^2_s/(16\pi^2f^2_S\Lambda^2)<<1$, due to the
factor of $1/(16\pi^2) $  which comes from the  loop momentum
integration for the diagrams  presented in Fig.1(b). Therefore,
one can safely neglect high order corrections   coming from
effective interaction Eq.\ref{lag}.

\section{Conclusion}

In summary, we consider the properties of scalar and pseudoscalar
glueballs in quark-gluon plasma. In the effective Lagrangian
approach, based on the low-energy QCD theorems, it is suggested
that scalar glueball remains massive above deconfinement
temperature. At the same time, pseudoscalar glueball changes its
properties in QGP in a drastic way. Indeed, this glueball become
massless at $T>T_c$ and therefore it can contribute strongly to
the bulk properties of QGP. We demonstrate that the disappearance
of pseudoscalar glueball mass above the deconfinement temperature
and its strong coupling to gluons gives the rise to the sign
change of the gluon condensate in the pure $SU(3)_c$ gauge theory
as observed in the lattice calculations at $T\approx T_c$. The
strong non-perturbative coupling of the glueball to the gluons
leads  to the conjecture that one might expect that the role of
very light pseudoscalar glueball in QGP must be quite similar to
the role played by the massless pion in nuclear matter below
deconfinement temperature. In spite of the
rather simple form of non-perturbative glueball-gluon interaction
and the neglect of high order non-perturbative and
perturbative interactions, as well as the possible
double-counting arising from the simultaneous consideration of hadronic and
partonic degrees of freedom, the possibility of the existence of very
light pseudoscalar glueball above $T_c$ is worth being considered in realizing
 the survival of the strong coupling between gluons
 via pseudoscalar glueball
  in the region of the temperature above $T_c$.
We like to mention on the lattice data in which bulk properties
of QGP do not change much with the inclusion of light quarks
\cite{lattice}. This may suggest that the appropriate extension
of our model to full QCD is possible~\cite{kochmin}.

\section{Acknowledgments}

We would like to thank A.Di Giacomo, A.E. Dorokhov, D.G. Pak,
S.-H. Lee and V.Vento
   for useful
discussions. This work was supported by Brain Pool program of
Korea Research Foundation through KOFST,  grant 042T-1-1. NK is
very grateful to the School of Physics and Astronomy of Seoul
National University for their warm hospitality during this work.

\end{document}